\documentclass[fleqn,usenatbib]{mnras}

\usepackage{newtxtext,newtxmath}

\usepackage[T1]{fontenc}

\DeclareRobustCommand{\VAN}[3]{#2}
\let\VANthebibliography\thebibliography
\def\thebibliography{\DeclareRobustCommand{\VAN}[3]{##3}\VANthebibliography}

\usepackage{graphicx}	
\usepackage{amsmath}	

\usepackage{amssymb}	
\usepackage{threeparttable}
\usepackage{multirow}
\usepackage{algorithm}
\usepackage{algpseudocode}
\usepackage{adjustbox}

\defcitealias{2015MNRAS.451.1728D}{DG15}

\title[Constraining galaxy geometry parameters]{Can we constrain galaxy geometry parameters using spatially integrated SED fitting?}

\author[Qiu et al.]{
Yisheng Qiu,$^{1}$\thanks{E-mail: yishengq@zju.edu.cn}
Xi Kang$^{1,2,3}$\thanks{E-mail: kangxi@zju.edu.cn}
Yu Luo,$^{2,3}$
\\
$^{1}$Institute for Astronomy, School of Physics, Zhejiang University, Hangzhou 310027, China \\
$^{2}$Purple Mountain Observatory, 10 Yuan Hua Road, Nanjing 210034, China \\
$^{3}$National Basic Science Data Center, Zhongguancun South 4th Street, Beijing 100190, China
}

\date{Accepted XXX. Received YYY; in original form ZZZ}

\pubyear{2015}

\begin{document}
\label{firstpage}
\pagerange{\pageref{firstpage}--\pageref{lastpage}}
\maketitle

\begin{abstract}
Sophisticated spectral energy distribution (SED) models describe dust attenuation and emission using geometry parameters. This treatment is natural since dust effects are driven by the underlying star-dust geometry in galaxies. An example is the \textsc{starduster} SED model, which divides a galaxy into a stellar disk, a stellar bulge, and a dust disk. This work utilises the \textsc{starduster} SED model to study the efficacy of inferring geometry parameters using spatially integrated SED fitting. Our method fits the SED model to mock photometry produced by combining a semi-analytic model with the same SED model. Our fitting results imply that the disk radius can be constrained, while the inclination angle, dust disk to stellar disk radius ratio, bulge radius and intrinsic bulge to total luminosity ratio are unconstrained, even though 21 filters from UV to FIR are used. We also study the impact of S/N, finding that the increase of S/N (up to 80) brings limited improvements to the results. We provide a detailed discussion to explain these findings, and point out the implications for models with more general geometry. 
\end{abstract}



\section{Introduction}
Spectral energy distribution (SED) fitting is a fundamental method to infer galaxy properties. These techniques have been applied to measure stellar masses, star formation rates, metallicities, dust masses, and dust attenuation curves \cite[e.g.][]{2017ApJ...837..170L,2018MNRAS.475.2891D,2020MNRAS.498.5581B,2022MNRAS.511..765Q}. What parameters can be inferred using SED fitting depends on how the SED is parameterised. A state-of-the-art SED code \citep[e.g][]{2019A&A...622A.103B,2020MNRAS.495..905R,2021ApJS..254...22J} describes a SED using parameters of star formation history (SFH), dust attenuation, dust emission and active galactic nucleus.
\par
An improvement of SED models could be the implementation of a more physical dust model. Currently, the dust attenuation curve in most SED models is described by a phenomenological model, e.g. a power law with additional parameters specifying the shape of the $2175 \, \text{\AA}$ bump. This method can be improved by taking into account more detailed star-dust geometry. For instance, the geometry can be parameterised using analytic density profiles. Given the geometry model, one can predict both dust attenuation and emission self-consistently using radiative transfer simulations \citep[e.g.][]{2020A&C....3100381C,2021ApJS..252...12N}. Following this approach, the SED is described by geometry parameters, and therefore one can infer these parameters by fitting the SED to observations. However, the application of SED models based on radiative transfer simulations \citep[e.g.][]{2013A&A...550A..74D,2019A&A...623A.143F} is not straightforward due to their high computational cost. Recently, this issue was partially overcome by \cite{2022ApJ...930...66Q}, who applied deep learning neural networks to accelerate the speed of a radiative transfer simulation. The SED model proposed by \cite{2022ApJ...930...66Q} describes dust attenuation and emission using geometry parameters. The model is named as \textsc{starduster}, and is publicly available on Github\footnote{\url{https://github.com/yqiuu/starduster}}.
\par
With the dust model in SED codes becoming more sophisticated, a natural question is whether the geometry parameters underlying the model can still be constrained using SED fitting. This work aims to address this question by fitting the \textsc{starduster} SED model to mock photometric data produced by coupling the same SED model with a semi-analytic model. This approach provides insight into the dependence of the SED on the geometry parameters. We then extend our discussion into more general geometry.
\par
This paper is organised as follows. We introduce the \textsc{starduster} SED model, the construction of the mock data, and the method of SED fitting in Section \ref{sec:method}. The fitting results are presented in Section \ref{sec:res}, followed by a detailed discussion in Section \ref{sec:discuss}. Finally, this work is summarised in Section \ref{sec:summary}.

\section{Methodology} \label{sec:method}
\subsection{The SED model} \label{sec:sed_model}
This work utilises the \textsc{starduster} SED model developed by \cite{2022ApJ...930...66Q}, which emulates the \textsc{skirt} radiative transfer simulation \citep{2011ApJS..196...22B,2015A&C.....9...20C,2020A&C....3100381C} using deep learning neural networks. The \textsc{starduster} model computes the dust-free SEDs by integrating the input SFH with the Flexible Stellar Population Synthesis library \citep{2009ApJ...699..486C,2010ApJ...712..833C}. For the library, we adopt a \cite{2003PASP..115..763C} initial mass function, and include nebular line and continuum emissions added by \cite{2017ApJ...840...44B}. The \textsc{starduster} model then reproduces the dust attenuation and emission calculations of the \textsc{skirt} simulation using neural networks. The underlying geometry of the model is described by a stellar disk, a stellar bulge, and a dust disk, with density profiles given by 
\begin{align}
    &\rho_\text{disk}(r, z) \propto \exp \left(- \frac{r}{r_\text{disk}} - \frac{|z|}{h_\text{disk}}\right), \label{eqn:rho_disk} \\ 
    &\rho_\text{bulge}(r, z) \propto \mathcal{S}_n \left(\frac{\sqrt{r^2 + z^2}}{r_\text{bulge}} \right), \label{eqn:rho_bulge} \\
    &\rho_\text{dust}(r, z) \propto \exp \left(- \frac{r}{q_\text{dust} r_\text{disk}} - \frac{|z|}{h_\text{dust}}\right), \label{eqn:rho_dust}
\end{align}
where $\mathcal{S}_n$ is the S\'ersic function. The model adopts $n = 4$. Moreover, we treat $r_\text{disk}$, $q_\text{dust}$, and $r_\text{bulge}$ as input parameters, and assume that $h_\text{dust} = h_\text{disk} = 0.1r_\text{disk}$. The reader is referred to \cite{2022ApJ...930...66Q} for a detailed description of the \textsc{starduster} SED model.

\begin{table*}
	\centering
	\caption{Full parameters of the \textsc{starduster} SED generative model.}
	\label{tab:params_raw}
	\begin{tabular}{llll} 
		\hline
		Symbol & Description & Range \\
		\hline
		$\theta$ & Inclination angle & 0 - 90 deg\\
		$r_\text{disk}$ & Stellar disk radius & $10^{-2.5} - 10^{2.5}$ kpc \\
		$r_\text{bulge}$ & Stellar bulge radius & $10^{-0.5} - 10^{1.5}$ kpc \\
		$q_\text{dust}$ & Dust disk radius to stellar disk radius (DTS) ratio & $10^{-0.7} - 10^{0.7}$ \\
		$\Sigma_\text{dust}$ & Dust surface density$^\text{a}$ & $10^{3.0} - 10^{7.5} \, \text{M}_\odot/\text{kpc}^2$ \\
		$l_\text{norm}$ & Intrinsic bolometric luminosity & $10^6 - 10^{14} \, \text{L}_\odot$ \\
		$\alpha_\text{B/T}$ & Intrinsic Bulge to total luminosity ratio & 0 - 1 \\
		$c^{\rm disk}_{i,k}$ & Luminosity fraction of the stellar disk in the $i$-th metallicity bin and the $k$-th stellar age bins$^\text{b}$ & 0 - 1  \\
		$c^{\rm bulge}_{i,k}$ & Luminosity fraction of the stellar bulge in the $i$-th metallicity bin and the $k$-th stellar age bin$^\text{b}$ & 0 - 1  \\
		\hline
	\end{tabular}
	\begin{tablenotes}
        \item $^\text{a}$ Dust surface density is defined by $\Sigma_\text{dust} = m_\text{dust} / (2\pi r^2_\text{dust})$.
        \item $^\text{b}$ There are 6 metallicity and 6 stellar age bins in the \textsc{starduster} model.
    \end{tablenotes}
\end{table*}

\begin{figure}
	\includegraphics[width=\columnwidth]{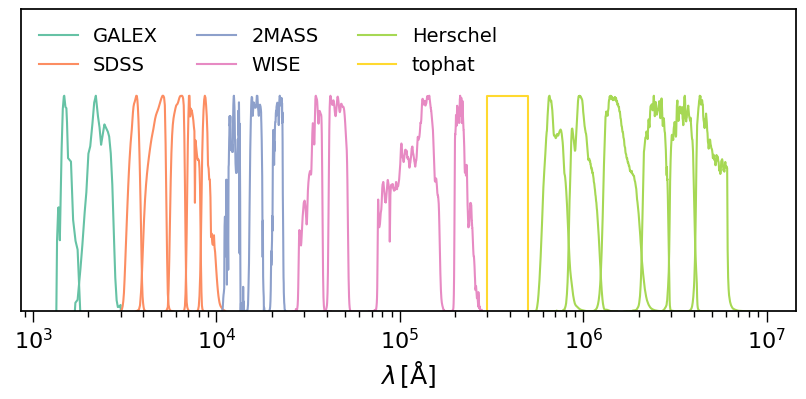}
    \caption{Adopted filters for the mock photometry.}
    \label{fig:filter}
\end{figure}

\begin{figure*}
	\includegraphics[width=.9\textwidth]{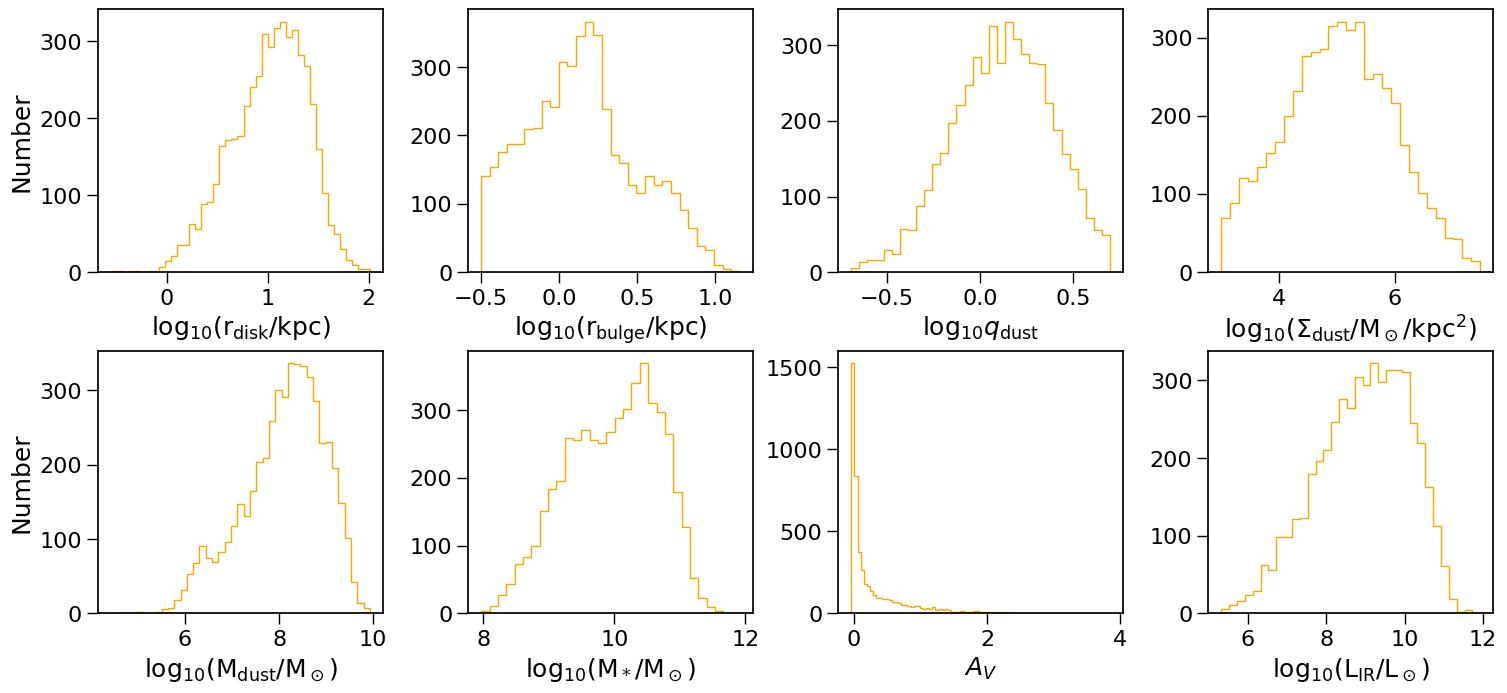}
    \caption{Upper panels: Distribution of stellar disk radius, bulge radius, DTS ratio, and dust surface density in the mock galaxy catalogue. Lower panels: Distribution of dust mass, stellar mass, V-band attenuation and total IR luminosity in the mock galaxy catalogue. All y-axes are in a linear scale. The construction of the catalogue is described in Section \ref{sec:mock}.}
    \label{fig:sample}
\end{figure*}

\subsection{Mock photometry} \label{sec:mock}
To construct a mock photometric catalogue, we combine the \textsc{starduster} SED model with a semi-analytic galaxy formation model. This work adopts the semi-analytic model presented by \cite{2016MNRAS.458..366L}, which is a modified version of the L-Galaxies model \citep{1996MNRAS.281..487K,2007MNRAS.375....2D,2011MNRAS.413..101G,2013MNRAS.428.1351G,2013MNRAS.434.1531F}. The model builds on the halo merger trees constructed from the Millennium Simulation \citep{2005Natur.435..629S}. Physics implemented in the model includes gas cooling, star formation, supernova feedback, metal production, bulge formation, and active galactic nuclei feedback. \cite{2016MNRAS.458..366L} verified that the model can reproduce the stellar, H$_1$, and H$_2$ mass functions at $z \sim 0$. The reader is referred to \cite{2011MNRAS.413..101G}, \cite{2013MNRAS.434.1531F}, and \cite{2016MNRAS.458..366L} for a detailed description of the model.
\par
The required galaxy parameters by the \textsc{starduster} SED model are listed in Table \ref{tab:params_raw}. In our semi-analytic model, the disk radius is proportional to the spin parameter and virial radius of the host halo. \cite{2013MNRAS.434.1531F} showed that this treatment leads to a median stellar surface density that is broadly consistent with the observations by \cite{2008AJ....136.2782L}. The bulge radius is predicted by the bulge formation model introduced in \cite{2011MNRAS.413..101G}, who demonstrated that their results agree with the observed half mass radius of early-type galaxies by \cite{2003MNRAS.343..978S}. The intrinsic bolometric luminosity, bulge to total luminosity ratio, and luminosity fractions are converted using the predicted disk mass, bulge mass, and SFH of the stellar disk and bulge. Additionally, we draw the inclination angle uniformly from a unit sphere.
\par
Dust properties are not predictions of the \cite{2016MNRAS.458..366L} semi-analytic model, and hence need additional modellings. First, we draw the logarithmic dust disk radius to stellar disk radius (DTS) ratio $\log_{10} q_\text{dust}$ from a normal distribution with $\mu = 0.15$ and $\sigma = 0.2$. This is based on the observations of nearby galaxies by \cite{2017A&A...605A..18C}, who estimated that the DTS ratio is $\sim 1 - 2$ depending on the wavelength in which the stellar disk radius is measured. Moreover, this work assumes that dust mass is proportional to cold gas mass \citep[e.g.][]{2016MNRAS.462.1057C}:
\begin{equation}
    m_\text{dust} = f_\text{dust} m_\text{cold}.
\end{equation}
We choose $f_\text{dust} = 0.1$. The 16-th and 84-th percentiles of the resulting dust surface density distribution are $10^4 \, \text{M}_\odot/\text{kpc}^2$ and $10^6 \, \text{M}_\odot/\text{kpc}^2$, which are comparable to observations of nearby galaxies \citep{2013ApJ...771...62K,2017A&A...605A..18C}.

\par
We compute the fluxes in 21 filters from UV to FIR for each galaxy. The adopted filters are shown in Figure \ref{fig:filter}. In addition to the commonly used filters, we also include a tophat filter at $40 \, \micron$. This is motivated by our previous study \citep{2022ApJ...930...66Q}, who found that the extra data could improve the measurement of bulge size. Moreover, we perturb the fluxes in all bands. We assume the same signal-to-noise (S/N) ratio in all bands, and adopt S/N values of 5, 10, 20, 40, and 80.
\par
We select model galaxies as follows. First, as listed in Table \ref{tab:params_raw}, each input parameter of the \textsc{starduster} SED has an allowed range, and we only include galaxies whose properties are within these ranges. In our model, a majority of low stellar mass galaxies have zero dust mass, which are excluded in this step. Most of these galaxies are satellites. The cold gas in these galaxies is removed by ram pressure stripping. Secondly, we only study galaxies at $z = 0$. The model produces millions of galaxies at $z = 0$, and it is unnecessary to include all of them. Instead, we randomly select 5,000 galaxies for the catalogue. Figure \ref{fig:sample} illustrates the distribution of some relevant properties in the mock catalogue.

\begin{table*}
	\centering
	\caption{Simplified parameters of the \textsc{starduster} SED fitting model.}
	\label{tab:params_fit}
	\begin{tabular}{llllll} 
		\hline
		Symbol & Description & Model$^\text{a}$ & Scale & Unit & Range \\
		\hline
		$\theta$ & Inclination angle && linear & deg & 0 - 90\\
		$r_\text{disk}$ & Stellar disk radius && log & kpc & -2.0 - 2.5\\
		$r_\text{bulge}$ & Stellar bulge radius && log & kpc & -0.5 - 1.5 \\
		$q_\text{dust}$ & Dust disk radius to stellar disk radius (DTS) ratio && log & - & -0.7 - 0.7 \\
		$\Sigma_\text{dust}$ & Dust surface density$^\text{b}$ && log & $\text{M}_\odot/\text{kpc}^2$ & 3.0 - 7.5 \\
		$l_\text{norm}$ & Intrinsic bolometric luminosity && log & $\text{L}_\odot$ & 6 - 14 \\
		$\alpha_\text{B/T}$ & Intrinsic Bulge to total luminosity ratio && linear & - & 0 - 1 \\
        \hline
		$a_1^\text{disk}$ & Transformed luminosity fraction for 1 Myr - 4 Myr & Discrete SFH & linear & - & 0 - 1 \\
		$a_2^\text{disk}$ & Transformed luminosity fraction for 4 Myr - 24 Myr & Discrete SFH & linear & - & 0 - 1 \\
		$a_3^\text{disk}$ & Transformed luminosity fraction for 24 Myr - 118 Myr & Discrete SFH & linear & - & 0 - 1 \\
		$a_4^\text{disk}$ & Transformed luminosity fraction for 118 Myr - 580 Myr & Discrete SFH & linear & - & 0 - 1 \\
		$a_5^\text{disk}$ & Transformed luminosity fraction for 580 Myr - 2 Gyr & Discrete SFH & linear & - & 0 - 1 \\
		$a_6^\text{disk} $ & Transformed luminosity fraction for 2Gyr - 14 Gyr & Discrete SFH & linear & - & 0 - 1 \\
		$\tau^\text{disk}$ & Star formation time scale & Exponentially Declining SFH & log & yr & 9.1 - 12.1 \\
		$t^\text{disk}_0$ & Star formation start time  & Exponentially Declining SFH  & log & yr & 6.0 - 10.1 \\
		$Z_\text{final}^\text{disk}$ & Maximum metallicity && log & $Z_\odot$ & -2.62 - 0.12 \\
	    $\xi^\text{disk}$ & Metallicity ratio && log & - &-3 - 0 \\
	    \hline
		$c^\text{bulge}$ & luminosity fraction for 580 Myr - 2 Gyr && linear & - & 0 - 1 \\
		$Z^\text{bulge}$ & Bulge metallicity && log & - & -2.62 - 0.12 \\
		\hline
	\end{tabular}
	\begin{tablenotes}
        \item $^\text{a}$ We adopt two different SFH models for the stellar disk as described in Section \ref{sec:fitting}.
        \item $^\text{b}$ Dust surface density is defined by $\Sigma_\text{dust} = m_\text{dust} / (2\pi r^2_\text{dust})$.
    \end{tablenotes}
\end{table*}

\subsection{SED fitting} \label{sec:fitting}
The application of the \textsc{starduster} model to SED-fitting requires a parameterisation of the SFH and metallicity history (MH). As listed in Table \ref{tab:params_raw}, the \textsc{starduster} model has 79 input parameters, which might be unsuitable for SED fitting. Our proposed model for SED fitting has 13 or 17 free parameters, which are described as follows. A summary of these parameters is given in Table \ref{tab:params_fit}.
\par
For the stellar disk, we adopt a discrete SFH, which uses 6 parameters to specify the luminosity fractions in 6 age bins. The bin edges are given in Table \ref{tab:params_fit}. Discrete SFH models are more flexible than traditional parametric models, which can reduce the bias of the inferred properties \citep{2019ApJ...876....3L,2020ApJ...904...33L}. We note that the luminosity fractions $c_i$ should satisfy
\begin{equation}
    \sum_i c_i = 1
\end{equation}
Following \cite{2022ApJ...930...66Q}, we use a transformation to deal with this constrain:
\begin{equation}
    c_i = \frac{\log(1 - a_i)}{\sum_i \log(1 - a_i)}. \label{eqn:simplex}
\end{equation}
The input parameters $a_i$ must be between 0 and 1. Equation \ref{eqn:simplex} transforms uniformly distributed samples into samples that follow a flat Dirichlet distribution.
\par
\begin{figure*}
	\includegraphics[width=.7\textwidth]{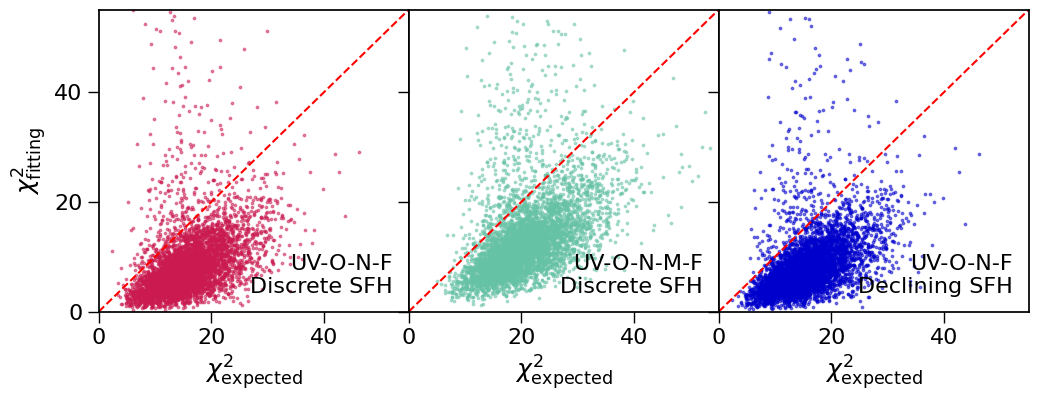}
    \caption{Relation between the best fitting and expected $\chi^2$. The expected $\chi^2$ is defined by the unperturbed and perturbed fluxes, and is calculated by replacing $f^\text{pred}_i$ in Equation \ref{eqn:chi2} with the unperturbed fluxes. These panels compare the results that include more filters or adopt a simpler SFH model. The two filter sets, i.e. UV-O-N-F and UV-O-N-M-F are described in Section \ref{sec:mock}. The discrete and declining SFH models are introduced in Section \ref{sec:fitting}.}
    \label{fig:chi2}
\end{figure*}

Alternatively, we propose an exponentially declining SFH model for the stellar disk. As will be shown in Figure \ref{fig:chi2}, the discrete SFH leads to overfitting. We therefore test whether the problem can be resolved by using a simpler model. The exponentially declining SFH as a function of stellar age is defined by
\begin{equation}
    \psi(t) = \begin{cases}
			e^{-t/\tau^\text{disk}}, & t < t^\text{disk}_0 \\
            0, & t \geq t^\text{disk}_0,
    \end{cases}
\end{equation}
where $\tau^\text{disk}$ and $t^\text{disk}_0$ are free parameters.
\par
For the MH of the stellar disk, we employ the model described in \cite{2020MNRAS.498.5581B} and \cite{2020MNRAS.495..905R}. This model assumes that the metallicity is a linear function of the cumulative SFH, which approximates the closed-box metallicity evolution. The metallicity in the $i$-th age bin is calculated by 
\begin{equation}
    Z_i = Z_\text{final} - (Z_\text{final} - Z_\text{min})(1 - \xi) \frac{\sum_{k}^{n=7 - i} m_k}{\sum_k^{n=6} m_k},
\end{equation}
where $m_k$ is the mass in the $k$-th age bin, and $Z_\text{min}=0.0024 \, Z_\odot$\footnote{We assume = $Z_\text{min}=0.019 \, Z_\odot$ in this work.} is the minimum metallicity of the adopted simple stellar population library. The free parameters in the model are $Z_\text{final}$ and $\xi$.
\par
For the stellar bulge, we assume that this component only consists of old stellar populations. Specifically, we restrict the presence of nonzero luminosity fractions to the oldest two age bins. Due to the normalisation, the bulge SFH is characterised by only one parameter. Additionally, the MH of the stellar bulge is assumed to be constant, which can be described by one parameter.
\par
The $\chi^2$ of the SED fitting is defined by
\begin{equation} \label{eqn:chi2}
    \chi^2 = \sum_\text{bands} \left( \frac{f_i^\text{mock} - f_i^\text{pred}}{\sigma} \right)^2,
\end{equation}
where $f_i^\text{mock}$ is the mock flux and $f_i^\text{pred}$ is the flux predicted by the SED model. A two-phase method is employed to minimise Equation \ref{eqn:chi2}. The aim of the first phase is exploration. We run a particle swarm optimiser \citep[e.g.][]{2012PhRvD..85l3008P} for 200 iterations. Using the minimum found in the first phase as the initial value, we subsequently adopt the L-BFGS-B algorithm \citep{l-bfgs-b} to polish the result. We find that minimising Equation \ref{eqn:chi2} is a challenging task. To ensure the convergence of the results, for each galaxy, we repeat the two-phase method three times, and choose the best-fitting parameters from all the runs.
\par
We consider the following variations in our SED fitting analysis:
\begin{enumerate}
    \item We fit our SED model to two filter sets, labelled as UV-O-N-M-F and UV-O-N-F. The former includes all 21 filters demonstrated in Figure \ref{fig:filter}. This filter set covers almost all wavelengths from UV to FIR, which is used to test whether the geometry parameters can be constrained in the ideal case. For the UV-O-N-F set, we exclude the fluxes from the WISE and tophat filters. In practice, MIR data are sensitive to dust grain composition, which is fixed in the \textsc{starduster} SED model. For both fitting processes, we assume a S/N of 20 and adopt the discrete SFH.
    \item We carry out a fitting process that uses the exponentially declining SFH for the stellar disk. As will be shown in Section \ref{sec:res}, the model based on the discrete SFH is overfitted. We test whether the overfitting can be reduced by using a simpler model. For this run, we choose the UV-O-N-F filter set and adopt a S/N of 20.
    \item The impact of S/N is also examined. Our SED model is fit to the mock data with S/N values of 5, 10, 20, 40, and 80. In each case, we use the UV-O-N-F filter set and assume the discrete SFH model.
\end{enumerate}

\begin{figure*}
	\includegraphics[width=\textwidth]{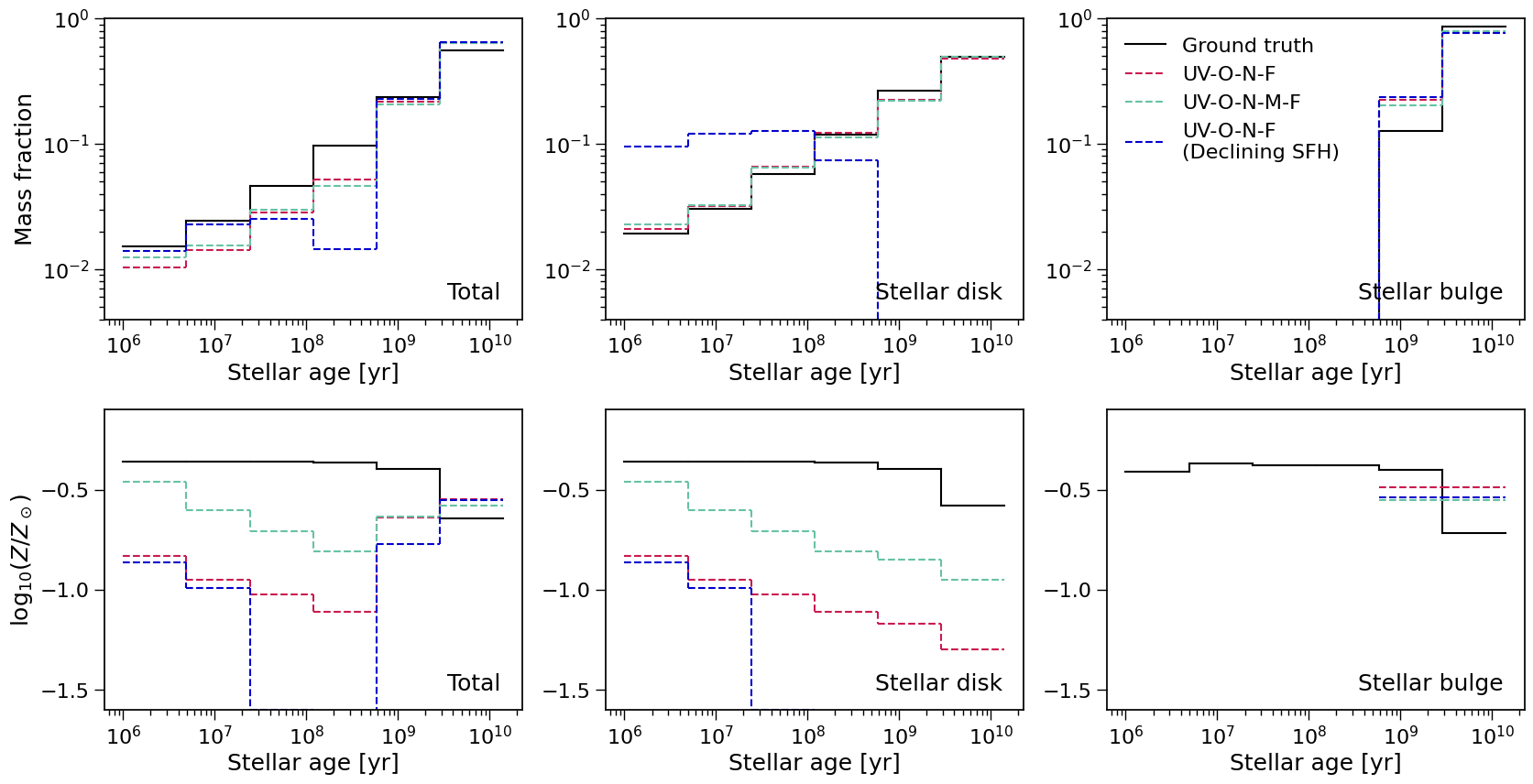}
    \caption{Recovery of SFHs and MHs. The upper and lower panels show the median mass fraction and metallicity as a function of stellar age respectively. Each panel compares the results that include more filters or adopt a simpler SFH model. The two filter sets, i.e. UV-O-N-F and UV-O-N-M-F are described in Section \ref{sec:mock}. The red and green line results are based on the discrete SFH, while the blue line results are based on the exponentially declining SFH. Both SFH models are described in Section \ref{sec:fitting}.}
    \label{fig:sfh}
\end{figure*}

\begin{figure*}
	\includegraphics[width=\textwidth]{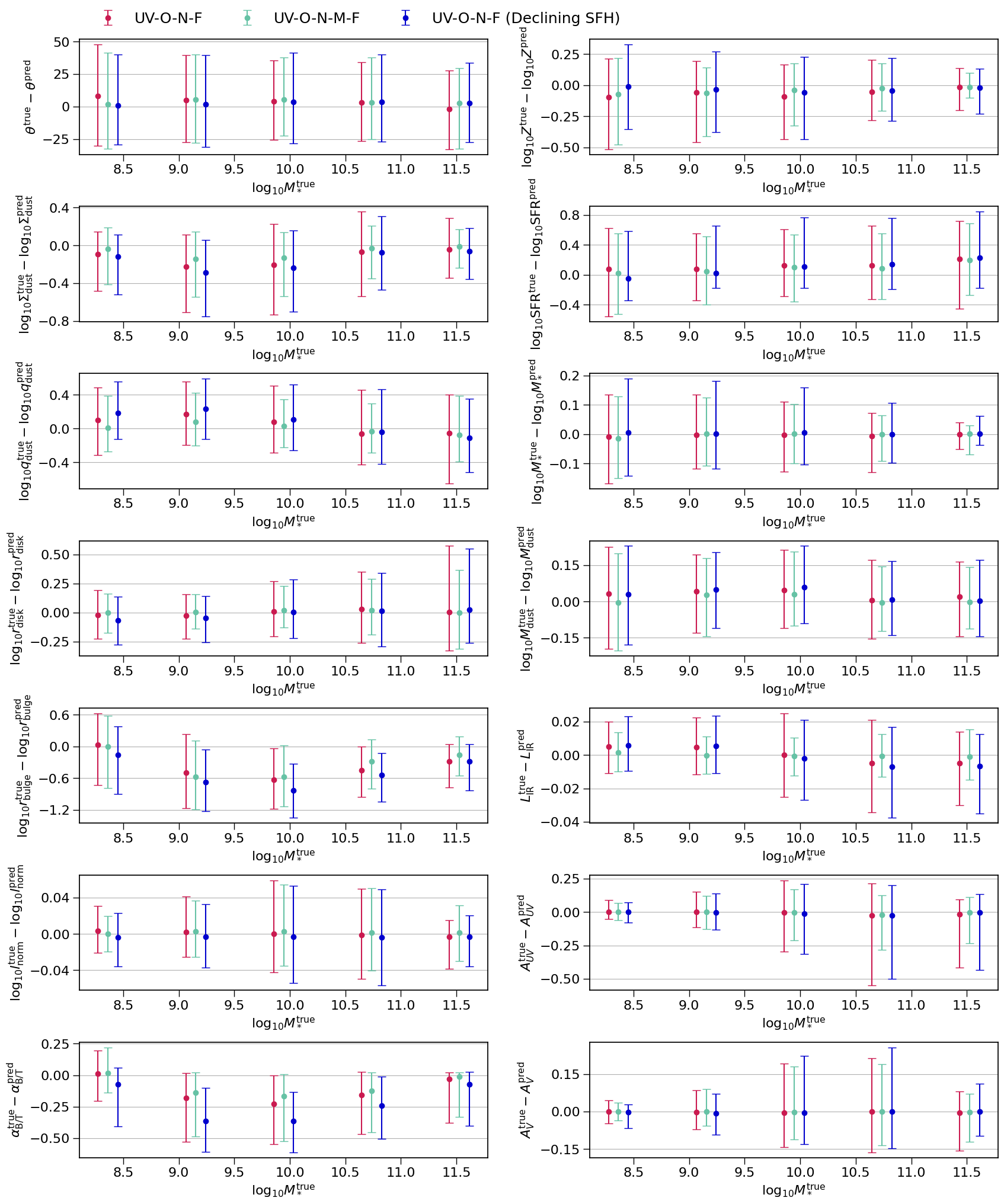}
    \caption{Distribution of the error between the ground truth and best fitting parameters as a function of stellar mass. Each panel compares the results that include more filters or adopt a simpler SFH model. We introduce the two filter sets, i.e. UV-O-N-F and UV-O-N-M-F in Section \ref{sec:mock}. The discrete and exponentially declining SFH models are described in Section \ref{sec:fitting}. The properties in the left panels are the direct input parameters of the SED model as listed in Table \ref{tab:params_fit}. In the right panels, from top to bottom, the properties are mass-weighted metallicity, star formation rate, stellar mass, dust mass, total infrared luminosity, UV band dust attenuation, and V band dust attenuation. The star formation rate timescale is 100 Myr.}
    \label{fig:params}
\end{figure*}

\begin{figure*}
	\includegraphics[width=\textwidth]{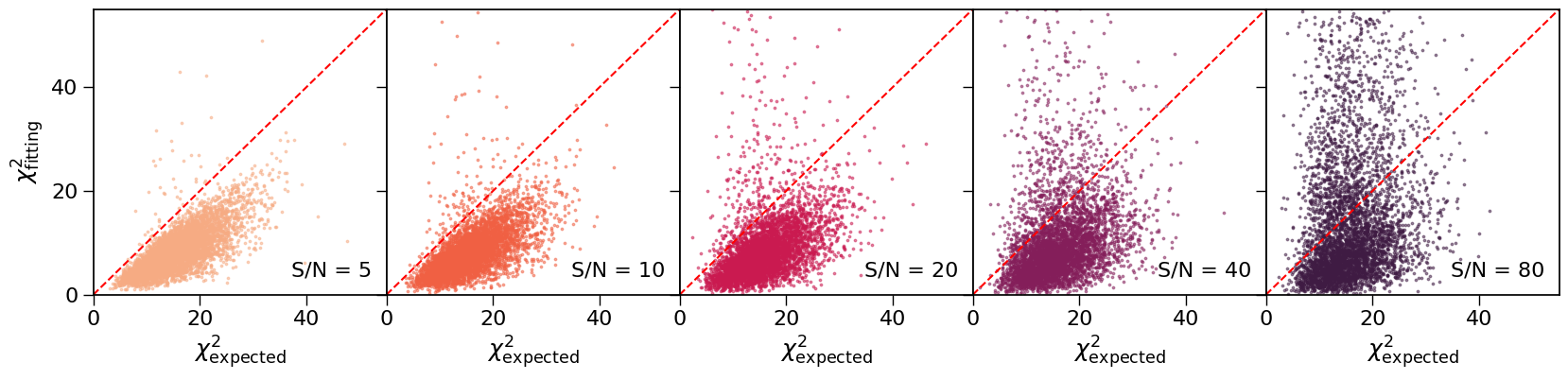}
    \caption{Relation between the best fitting and expected $\chi^2$ as a function of S/N. The expected $\chi^2$ is defined by the unperturbed and perturbed fluxes, and is calculated by replacing $f^\text{pred}_i$ in Equation \ref{eqn:chi2} with the unperturbed fluxes. These results are based on the UV-O-N-F filter set (see Section \ref{sec:mock}) and the discrete SFH model (see Section \ref{sec:fitting}).}
    \label{fig:chi2_sn}
\end{figure*}

\begin{figure*}
	\includegraphics[width=\textwidth]{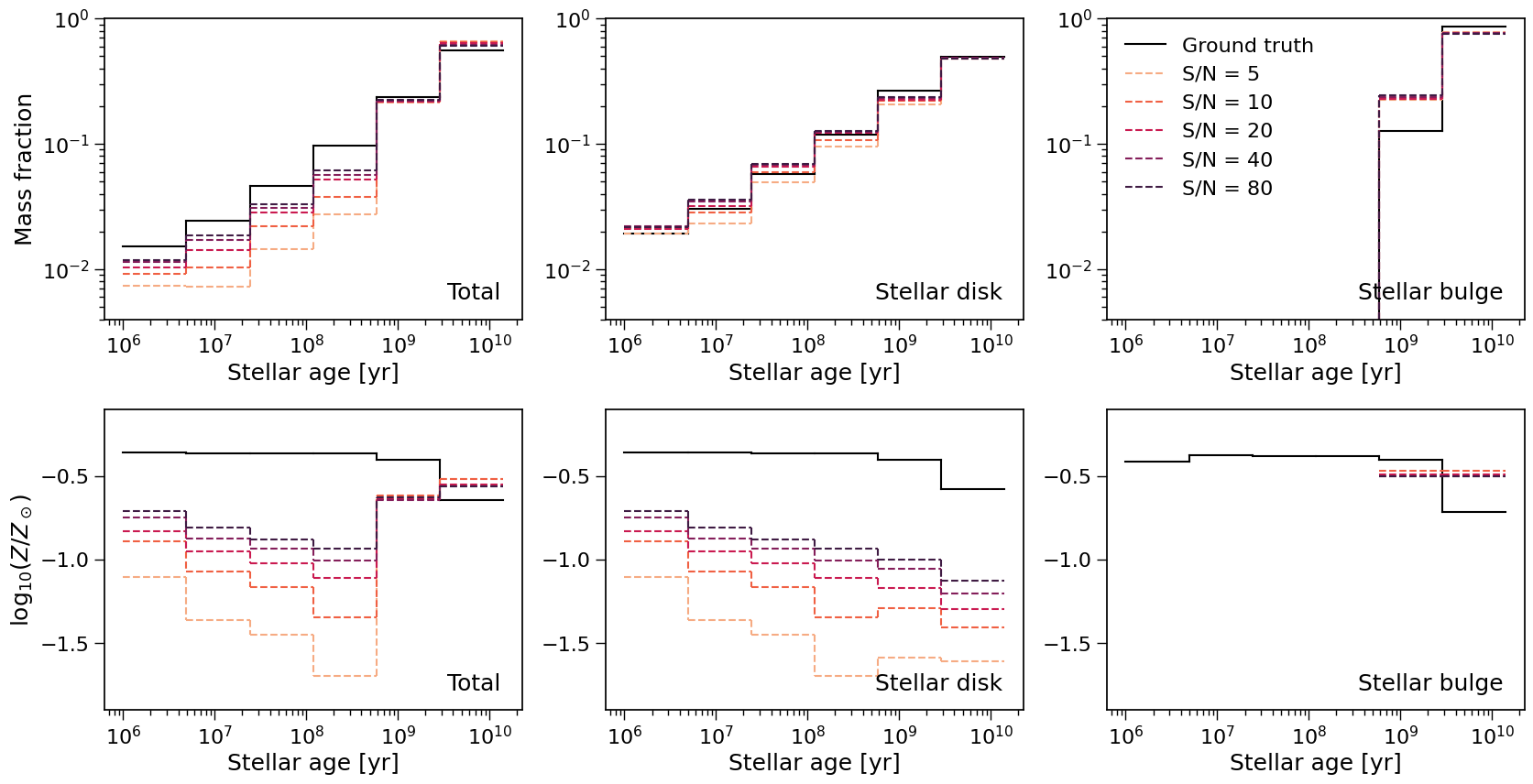}
    \caption{Recovery of SFHs and MHs as a function of S/N. The upper and lower panels show the median mass fraction and metallicity as a function of stellar age respectively. These results are based on the UV-O-N-F filter set (see Section \ref{sec:mock}) and the discrete SFH model (see Section \ref{sec:fitting}).}
    \label{fig:sfh_sn}
\end{figure*}

\begin{figure*}
	\includegraphics[width=\textwidth]{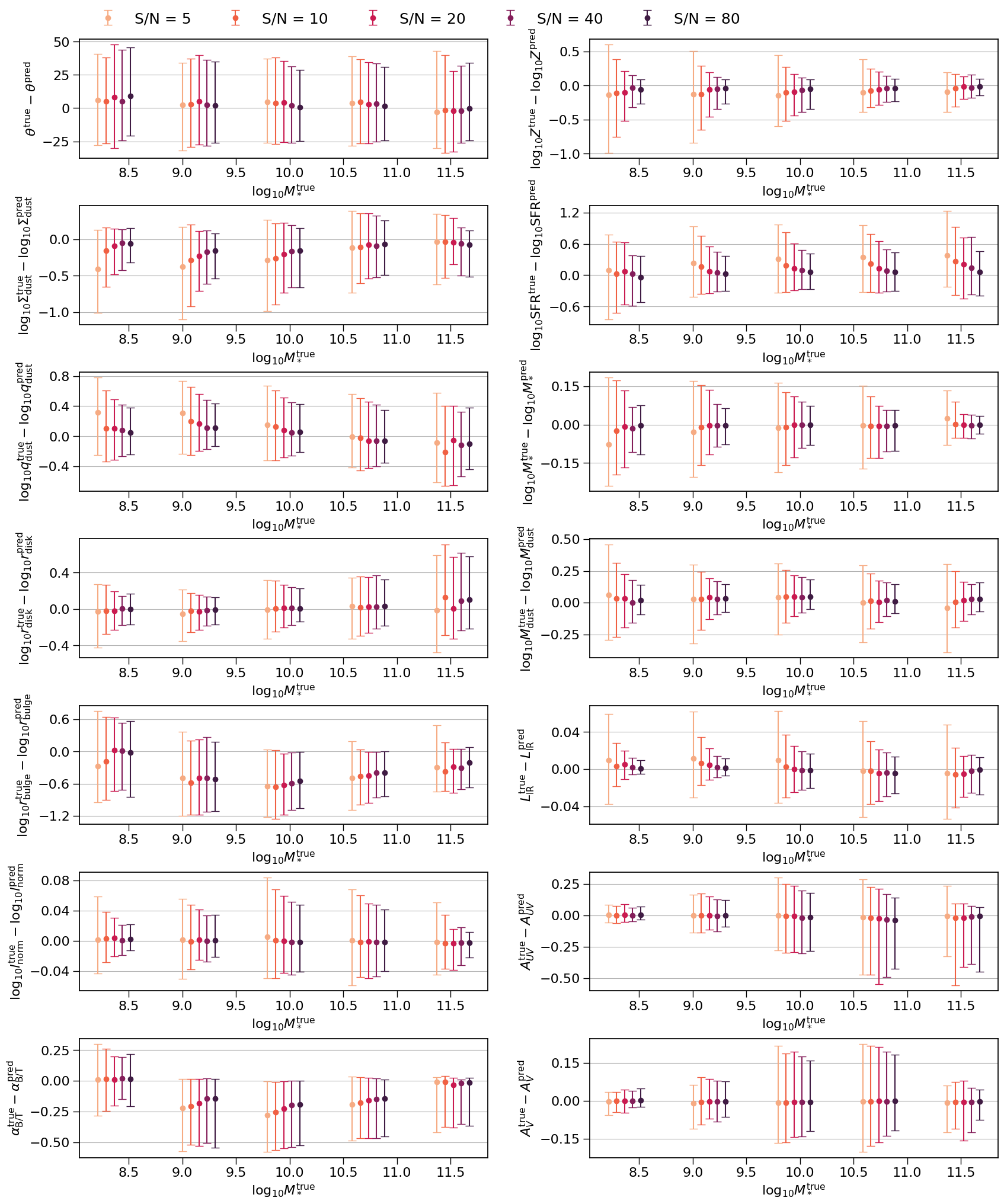}
    \caption{Distribution of the error between the ground truth and best fitting parameters as a function of stellar mass. Each panel compares the results for different S/N values. These results are based on the UV-O-N-F filter set (see Section \ref{sec:mock}) and the discrete SFH model (see Section \ref{sec:fitting}). The properties in the left panels are the direct input parameters of the SED model as listed in Table \ref{tab:params_fit}. In the right panels, from top to bottom, the properties are mass-weighted metallicity, star formation rate, stellar mass, dust mass, total infrared luminosity, UV band dust attenuation, and V band dust attenuation. The star formation rate timescale is 100 Myr.}
    \label{fig:params_sn}
\end{figure*}

\section{Results} \label{sec:res}
\subsection{Goodness of fit}
We start by examining the goodness of fit. Figure \ref{fig:chi2} compares the best-fitting and expected $\chi^2$ values. The expected values are calculated by replacing $f^\text{pred}_i$ in Equation \ref{eqn:chi2} with the unperturbed fluxes. As shown in Figure \ref{fig:chi2}, the best-fitting $\chi^2$ values are generally lower than the expected ones. While this confirms the convergence of our fitting algorithm, it also indicates overfitting. As illustrated in the middle and left panels of Figure \ref{fig:chi2}, including more filters or using a simpler model cannot resolve the issue. An implication is that the parameter space of our SED has strong degeneracies. Despite the problem, as will be shown below, we can still recover some of the galaxy properties.

\subsection{Recovery of SFHs and MHs}
The upper panels of Figure \ref{fig:sfh} compare the inferred SFHs with the input mock data. When using the discrete SFH model, we can recover the median mass fractions of the stellar disk in all age bins and those of the stellar bulge in the oldest age bin. However, for the total SFH, consistent results are only found in the 580 Myr $ < t <$ 14 Gyr age bins. An implication is that the bulge to total luminosity ratio is unconstrained. When using the exponentially declining SFH, we fail to recover the disk SFH. However, for the total SFH, except the 24 Myr $ < t <$ 580 Myr age bins, the inferred median mass fractions are consistent with the input mock data.
\par
The bottom panels of Figure \ref{fig:sfh} demonstrate the recovery of MHs. In all cases, we cannot obtain consistent results. We note that our proposed MH model in Section \ref{sec:fitting} depends on the SFH. Consequently, if the SFH cannot be well recovered, we will fail to obtain consistent MH. Despite this discrepancy, as shown in Figure \ref{fig:params}, we call still constrain the mass-weighted metallicity.

\subsection{Recovery of galaxy parameters}
In Figure \ref{fig:params}, we show the error of some recovered galaxy properties as a function of stellar mass. The error is defined by the difference between the truth and best-fitting values for each galaxy. We first focus on the results based on the UV-O-N-F filter set and the discrete SFH, corresponding to the red circles with error bars. The errors in the derived disk radius range from -0.2 to 0.25 dex in the low mass end and from -0.25 to 0.5 dex in the high mass end. This implies that the disk radius can be constrained, given that the input disk radius varies over 2 orders of magnitude as shown in Figure \ref{fig:sample}. In contrast, the errors in the inferred DTS ratio range from -0.3 to 0.4 dex, which is greater than the intrinsic scatter, i.e. 0.2 dex. Additionally, this property is also systematically overestimated in the high stellar mass end. Hence, the DTS ratio is unconstrained. Similarly, the errors in the inferred inclination angle are comparable with its allowed range, implying that this quantity is also unconstrained. Moreover, the best-fitting bulge radii and intrinsic bulge to total luminosity ratios show large systematic offsets. Therefore, we fail to infer these properties using SED fitting.
\par
When including more data in MIR to FIR wavelengths, as illustrated in Figure \ref{fig:params}, the fitting results are only slightly improved. Using the same SED model, \cite{2022ApJ...930...66Q} found that the measurement of the bulge radius can be improved when including data at $\sim 40 \, \micron$ in the SED fitting of two observed galaxies. However, our results in Figure \ref{fig:params} indicate that the $40 \, \micron$ flux in general is insufficient to recover the bulge radius. In addition, the other unconstrained parameters, i.e. the inclination angle, DTS ratio, bulge radius and intrinsic bulge to total luminosity ratio remain unconstrained when including more data.
\par
Figure \ref{fig:params} also illustrates the fitting results based on the exponentially declining SFH. In general, these results are similar to those based on the discrete SFH. A difference is that the inferred stellar mass and star formation rate show more bias. The reason is that the exponentially declining SFH model is not flexible enough to fit the mock data, which was discussed in great detail in previous studies \citep{2019ApJ...873...44C,2019ApJ...876....3L,2020ApJ...904...33L}.

\subsection{Trends with S/N}
We demonstrate the impact of S/N on the SED fitting results in Figures \ref{fig:chi2_sn}, \ref{fig:sfh_sn}, and \ref{fig:params_sn}. These results are based on the UV-O-N-F filter set and the discrete SFH. In Figure \ref{fig:chi2_sn}, we find that increasing the S/N cannot resolve the overfitting problem. Particularly, with the increase of the S/N, many results show a substantially large $\chi^2$ value, meaning that the SED model cannot fit the mock data. We compare the inferred SFHs and MHs at different S/N values in Figure \ref{fig:sfh_sn}. In general, higher S/N values lead to more consistent results. However, the improvement becomes less significant when the S/N is increased from 20 to 80. A similar trend is found in the recovery of galaxy parameters as illustrated in Figure \ref{fig:params_sn}. In particular, varying the S/N has no impact on the inferred inclination angle, bulge radius and intrinsic bulge to total luminosity ratio, which are all unconstrained parameters. All these findings imply that high S/N data cannot break the degeneracies in the SED model. In other words, the improvements from high S/N data are limited. A similar conclusion was obtained by \cite{2019ApJ...876....3L}, who found that the chosen prior of the SFH can have a larger impact on the fitting results than the photometric noise.

\begin{figure*}
	\includegraphics[width=\textwidth]{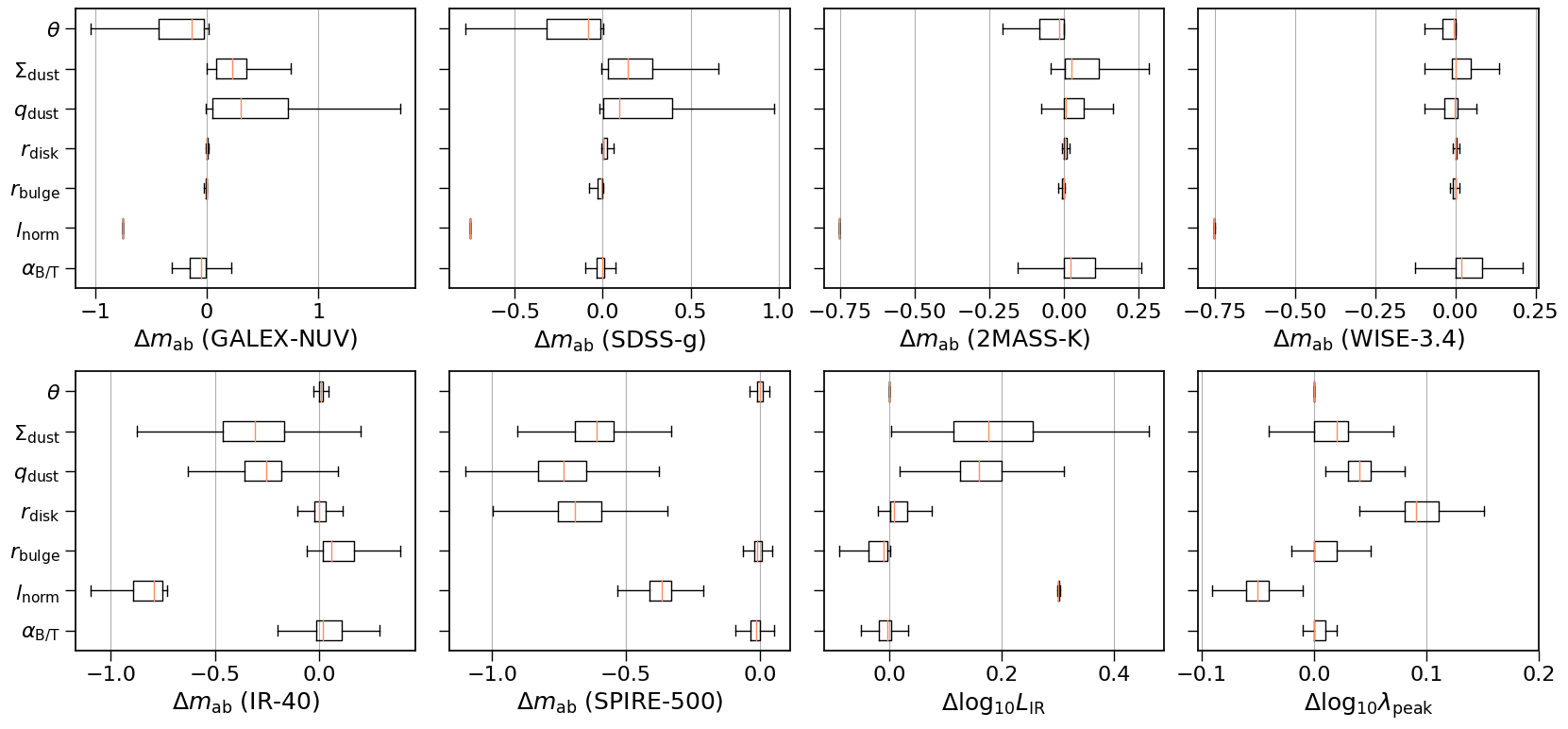} 
    \caption{Dependence of the SED on the input parameters. Each panel shows the variance of the AB magnitude in a filter when the input SED parameters are increased by a factor of 2, except the inclination angle, which is increased by 15 deg. In addition, the third and fourth panels in the bottom show the variance of the total infrared luminosity and the peak of the FIR emission respectively. These results are based on the mock galaxy catalogue described in Section \ref{sec:mock}. The definition of the varied parameters can be found in Table \ref{tab:params_fit}.}
    \label{fig:variance}
\end{figure*}

\section{Discussion} \label{sec:discuss}
From our fitting results presented in the previous section, we conclude that the disk radius can be constrained, while the DTS ratio, bulge radius, intrinsic bulge to total luminosity ratio, and inclination angle are unconstrained or poorly constrained. To interpret the results, we need to understand how the SED depends on these parameters. In Figure \ref{fig:variance}, we increase the input SED parameters by a factor of 2, and illustrate the resulting variance in some selected bands. The inclination angle is not in a logarithmic scale and is increased by 15 deg instead. The shape of the FIR spectrum can be characterised more intuitively using the total infrared luminosity and the peak wavelength of the FIR emission. Hence, we also show the variance of these two quantities in Figure \ref{fig:variance}.

\subsection{Disk radius} \label{sec:radius}
Figure \ref{fig:variance} illustrates that increasing the disk radius (at fixed DTS ratio) has no impact on the UV to optical SED but leads to the largest shift in the peak of the FIR emission. Besides the disk radius, the peak of the FIR emission is also sensitive to the intrinsic bolometric luminosity and the DTS ratio. The former gives the overall normalisation of the SED, and can be constrained by the UV to NIR fluxes. The latter has a small allowed range, and therefore cannot substantially change the absolute value of the peak wavelength. In other words, the degeneracies between the disk radius and these parameters are insignificant. Hence, the disk radius can be constrained by the peak of the FIR emission.
\par
The dependence of the SED on the disk radius found in this work was explained in the radiative transfer study by \cite{2001ApJ...551..277M}. They suggested that varying the physical size of a dusty system with all the other quantities fixed does not change the total optical depth, and hence has no effect on the UV to optical spectrum. \cite{2001ApJ...551..277M} also pointed out that the dust temperature depends on the ratio of the absorbed energy to the dust mass. In our model, increasing the disk radius at fixed dust surface density and DTS ratio results in higher dust mass and therefore lower dust temperature, which shifts the peak of the FIR emission to longer wavelengths.

\begin{figure*}
	\includegraphics[width=.9\textwidth]{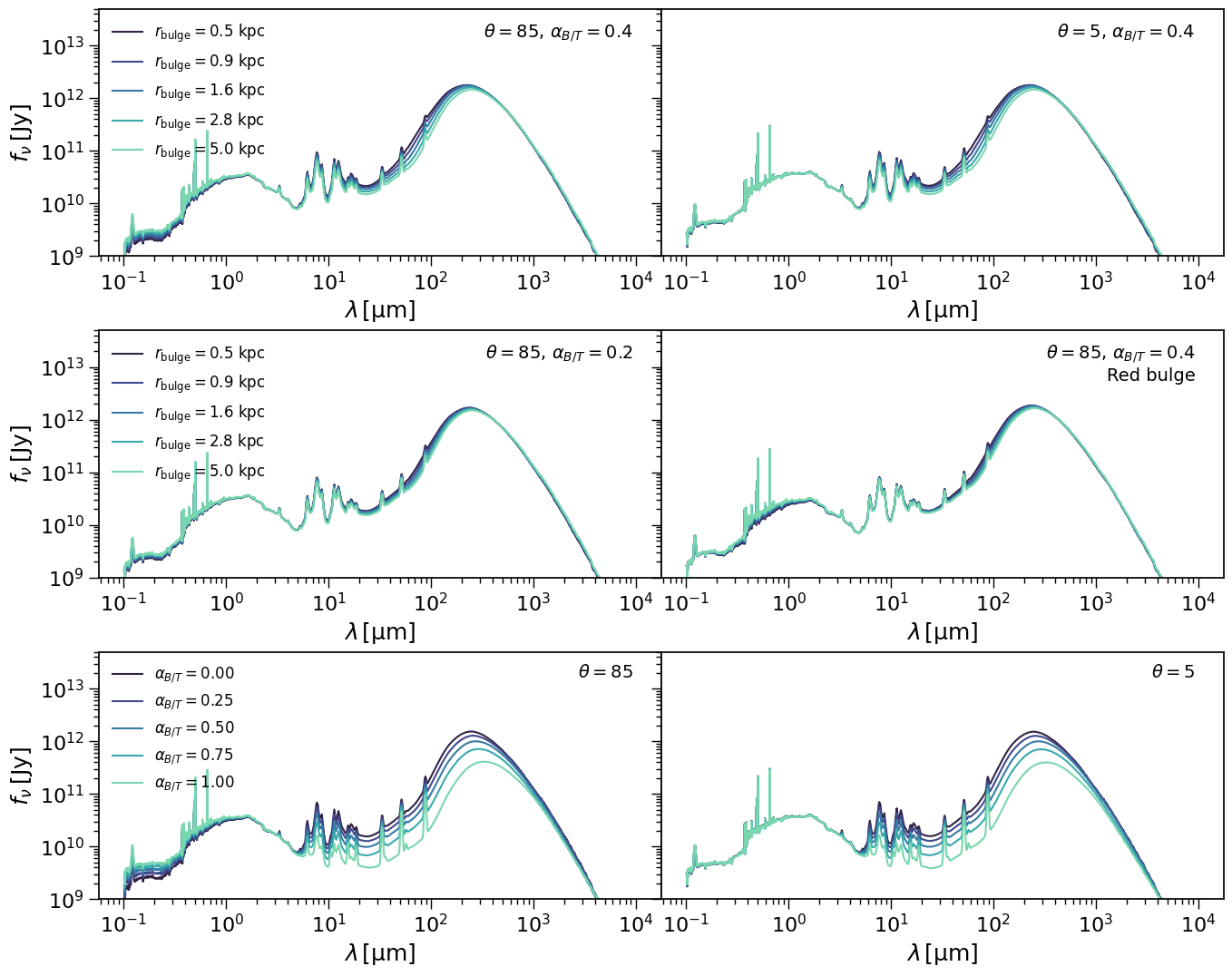}
    \caption{Upper and middle panels: dependence of the SED on the bulge radius at different inclination angles, intrinsic bulge to total luminosity ratios, and stellar populations. Lower panels: dependence of the SED on the intrinsic bulge to total luminosity ratio. The base SED and adopts $\Sigma_\text{dust} = 10^5 \, \text{M}_\odot / \text{kpc}^2$, $r_\text{disk} = 5 \, \text{kpc}$, $q_\text{dust} = 1$, $r_\text{buge} = 1 \, \text{kpc}$, and $l_\text{norm} = 10^9 \, \text{L}_\odot$. The SFH is assumed to be constant from 1 Myr to 14 Gyr, with $Z = 0.1 Z_\odot$. For the red bulge variant, the stellar population from 580 Myr to 14 Gyr is in the bulge, with the overall SFH kept constant.}
    \label{fig:bulge}
\end{figure*}

\begin{figure*}
	\includegraphics[width=.7\textwidth]{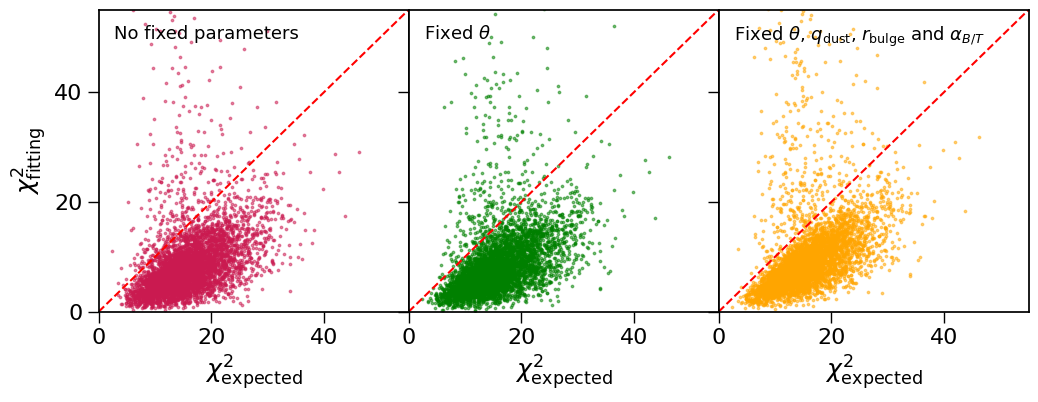}
    \caption{Relation between the best fitting and expected $\chi^2$. The expected $\chi^2$ is defined by the unperturbed and perturbed fluxes, and is calculated by replacing $f^\text{pred}_i$ in Equation \ref{eqn:chi2} with the unperturbed fluxes. These panels compare the results when certain parameters are fixed in the fitting. These results are based on the UV-O-N-F filter set (see Section \ref{sec:mock}) and the discrete SFH model (see Section \ref{sec:fitting}).}
    \label{fig:chi2_fix}
\end{figure*}

\begin{figure*}
	\includegraphics[width=\textwidth]{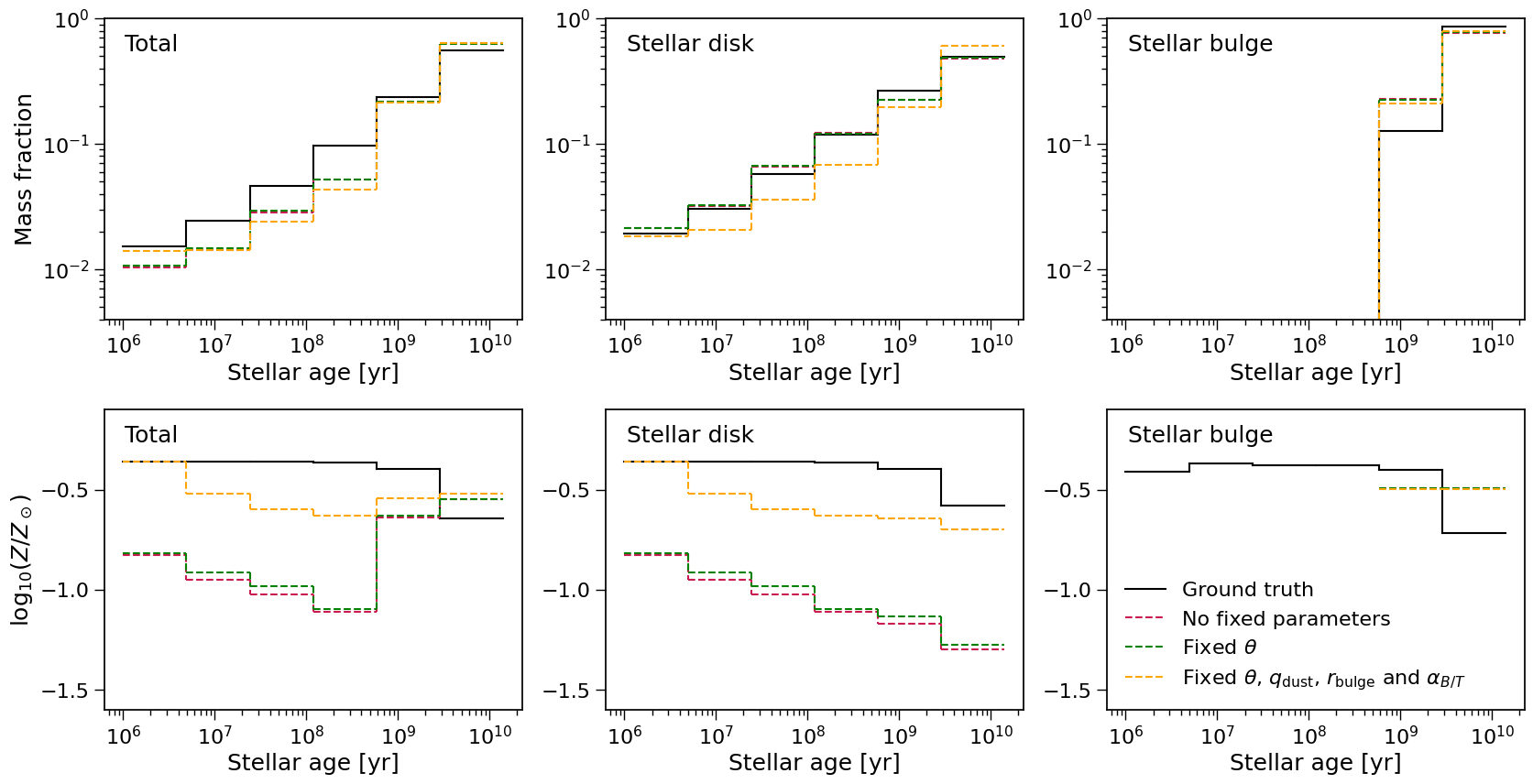}
    \caption{Impacts of fixing certain parameters in the fitting on the inferred SFHs and MHs. The upper and lower panels show the median mass fraction and metallicity as a function of stellar age respectively. Each panel compares the results when certain parameters are fixed in the fitting. These results are based on the UV-O-N-F filter set (see Section \ref{sec:mock}) and the discrete SFH model (see Section \ref{sec:fitting}). For the mass fractions, the results with no parameters fixed overlap those with the inclination angle fixed.}
    \label{fig:sfh_fix}
\end{figure*}

\begin{figure*}
	\includegraphics[width=\textwidth]{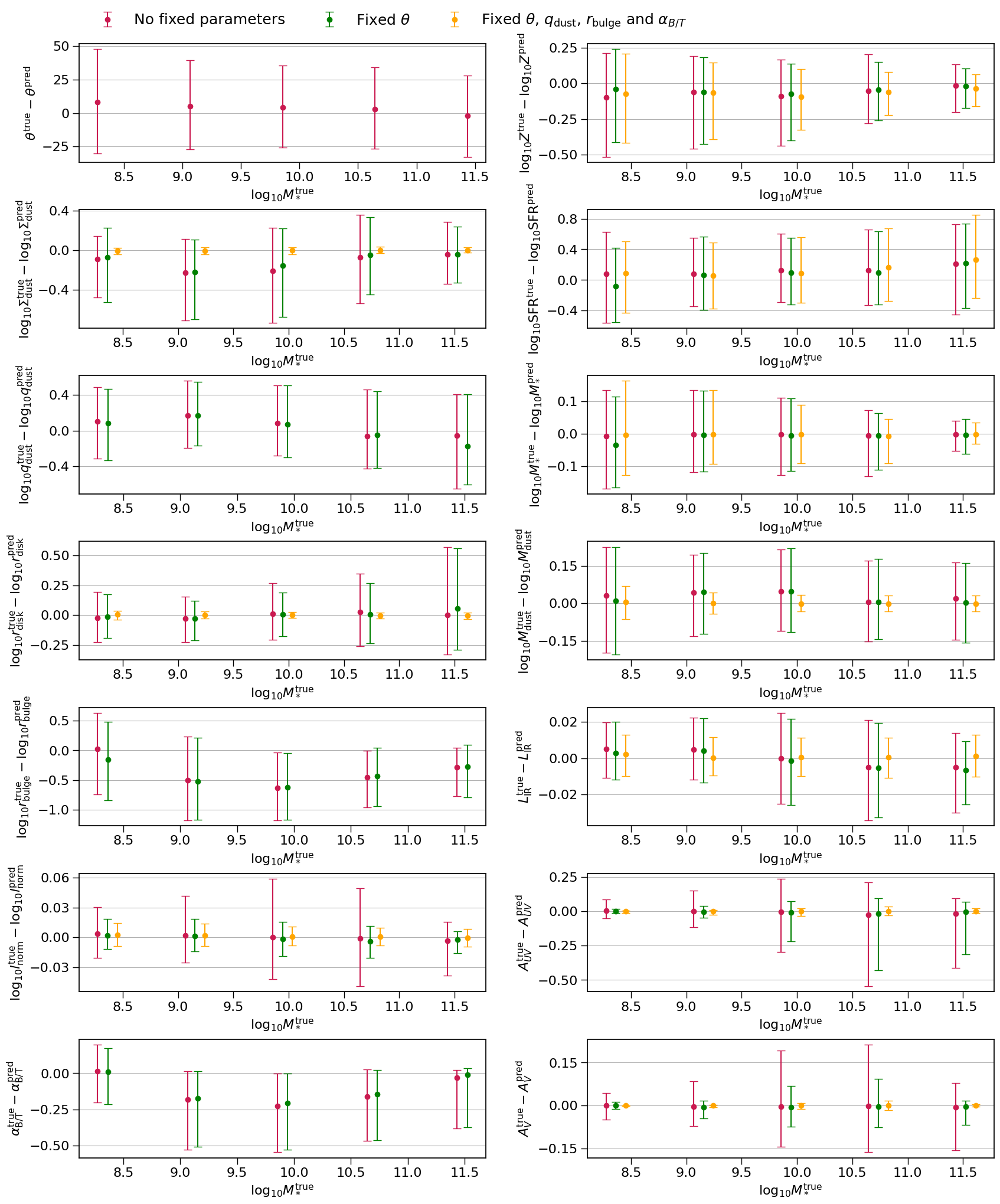}
	\caption{Impacts of fixing certain parameters in the fitting on the inferred galaxy properties. Each panel shows the error between the ground truth and best fitting parameters as a function of stellar mass. No data points will be plotted if the property is fixed. These results are based on the UV-O-N-F filter set (see Section \ref{sec:mock}) and the discrete SFH model (see Section \ref{sec:fitting}). The properties in the left panels are the direct input parameters of the SED model as listed in Table \ref{tab:params_fit}. In the right panels, from top to bottom, the properties are mass-weighted metallicity, star formation rate, stellar mass, dust mass, total infrared luminosity, UV band dust attenuation, and V band dust attenuation. The star formation rate timescale is 100 Myr.}
    \label{fig:params_fix}
\end{figure*}

\subsection{Dust surface density and DTS ratio} \label{sec:dts}
Figure \ref{fig:variance} shows that varying the dust surface density and the DTS ratio influences the UV to FIR fluxes in a similar way. However, only the dust surface density is constrained. A potential reason is that the allowed range of dust surface density is much larger than the DTS ratio. As listed in Table \ref{tab:params_fit}, the dust surface density can span over 4 orders of magnitude, while $\log_{10} q_{\rm dust}$ is only allowed to vary from -0.7 to 0.7. Hence, the dust surface density is the primary factor that determines the amount of dust attenuation, while the DTS ratio is a secondary term.

\subsection{Bulge radius and intrinsic bulge to total luminosity ratio} \label{sec:bulge}
Our results suggest that the bulge radius and the intrinsic bulge to total luminosity ratio cannot be constrained. To explain the reason, we need to understand how the SED depends on these quantities. If we assume that the stellar disk and bulge have the same stellar population, varying $r_\text{bulge}$ and $\alpha_\text{B/T}$ will similarly impact the SED. This point is demonstrated in Figure \ref{fig:bulge}. Decreasing the bulge radius or the intrinsic bulge to total luminosity ratio increases the fraction of light that is obscured by the dust disk. This effect reddens the UV to optical colour for highly inclined galaxies but is negligible for face-on galaxies. In MIR to FIR wavelengths, the decrease of the bulge radius or the intrinsic bulge to total luminosity ratio increases the absorbed energy and therefore the dust temperature. This slightly shifts the peak of the FIR emission to shorter wavelengths and increases the MIR to FIR fluxes. A similar effect of the intrinsic bulge to total luminosity ratio on the FIR spectrum was found by \cite{2011A&A...527A.109P}. In addition, when the stellar disk and bulge have different stellar populations, varying $\alpha_\text{B/T}$ can influence the NIR flux as shown in Figure \ref{fig:variance}, which, however, can be mimicked by the SFH parameters.
\par
We further point out two factors that weaken the dependence of the SED on the bulge radius. They are illustrated in the middle panels of Figure \ref{fig:bulge}. First, the effect of varying the bulge radius is only significant for large $\alpha_\text{B/T}$. Secondly, the dust effect of the bulge will be less pronounced if the bulge is dominated by old stellar populations.
\par
In short, varying the bulge radius and intrinsic bulge to total luminosity ratio only has moderate effects on the UV to optical and MIR to FIR fluxes, and there is a strong degeneracy between both parameters. Therefore, these parameters cannot be constrained using SED fitting.

\subsection{Inclination angle}
The inclination angle is also an unconstrained parameter. A potential reason is that the inclination angle is irrelevant to the dust emission spectrum but degenerates with all the other parameters in the UV to optical regime.
\par
Since the inclination angle can be measured using galaxy images, it is instructive to check the fitting results with this parameter fixed. Such results are presented in Figures \ref{fig:chi2_fix}, \ref{fig:sfh_fix}, and \ref{fig:params_fix}. A major improvement is that the $1\sigma$ offsets of the UV and V band attenuation are reduced by $\sim 0.1$ mag, while the estimations of the other galaxy parameters and the SFHs are unaffected. The impact of the inclination on measuring dust attenuation was also highlighted by \cite{2022ApJ...931...53D}, who carried out a SED-fitting analysis on 151 disk dominated galaxies using an inclination-dependent model.

\subsection{More general geometry}
Our discussion above can be generalised to different geometry models. Galaxy geometry is more complex and diverse in reality, which in general could not be described by the density profiles introduced in Section \ref{sec:sed_model}. However, the spatial distribution of stars and dust grains are correlated, since dust grains are produced by supernova and stellar winds. Accordingly, the star-dust geometry could be described by a common characteristic scale, with higher order terms quantifying the differences between the stellar and dust disks. Our analysis in Section \ref{sec:radius} implies that the common characteristic scale should have limited effects on the UV to optical spectrum but dramatically influence the peak of the FIR emission. Therefore, for a SED model with more complex geometry, we should be able to constrain the characteristic scale using FIR data. For the higher order terms, an analogue is the DTS ratio in our SED model. As illustrated in Figure \ref{fig:variance}, this parameter can impact all UV to FIR fluxes. However, as discussed in Section \ref{sec:dts}, the DTS ratio cannot be constrained due to its small allowed range. For the same reason, we expect that spatially integrated SED fitting cannot constrain the differences between the distribution of stars and dust in general. In addition, our discussion in Section \ref{sec:bulge} suggests that we are also unable to constrain bulge related properties.

\subsection{Improvement}
Our analysis implies that not all geometry parameters can be constrained using the spatially integrated SED, and a potential improvement is to combine SED fitting with image analysis tools. Such an idea was presented by \cite{2011A&A...527A.109P} and \cite{2022MNRAS.513.2985R}. A typical profile fitting tool \cite[e.g.][]{2010AJ....139.2097P} can estimate inclination angle, disk size, bulge size, and bulge to total luminosity ratio. One could also extract dust disk size from FIR images. These parameters are also inputs of the \textsc{starduster} SED model. Hence, we can fit the SED and the light profile jointly.
\par
A question is what can be improved when the unconstrained parameters in our SED model are determined using the extra information from galaxy images. To demonstrate, we perform an additional SED fitting process with all the unconstrained parameters fixed. The fitting results are shown in Figures \ref{fig:chi2_fix}, \ref{fig:sfh_fix}, and \ref{fig:params_fix}. First, the errors in the inferred dust surface density and the stellar disk radius are reduced substantially, and hence we obtain better estimations of the dust mass and dust attenuation. This is expected since fixing the unconstrained parameters removes the degeneracies between the geometry parameters. Secondly, we find no improvements in the inferred SFHs as illustrated in Figure \ref{fig:sfh_fix}. In particular, the recovered mass fractions of the disk become less consistent than the results with no parameters fixed. A potential reason is that there are degeneracies between the SFH parameters, which was pointed out by \cite{2019ApJ...876....3L}. The degeneracies could also explain the overfitting in Figure \ref{fig:chi2_fix}. In addition, the improvements in the inferred stellar mass, SFR, and mass-weighted metallicity are insignificant. The uncertainties of these quantities are mainly from the inferred SFH.

\section{Summary} \label{sec:summary}
This work examines the efficacy of incorporating detailed geometry models into spatially integrated SED fitting. Specifically, we investigate whether the geometry parameters in the \textsc{starduster} SED model can be constrained using SED fitting. The fitting data are produced by coupling the same SED model with a semi-analytic model. Our findings are summarised as follows:
\begin{itemize}
    \item We identify a correlation between the disk radius and the peak of the dust emission at fixed DTS ratio. The disk radius can be constrained using this feature. For the same reason, we suggest that the characteristic scale of the dust and stellar disk can also be constrained using SED fitting in more general geometry.
    \item The DTS ratio can impact both dust attenuation and emission, and has a degeneracy with the dust surface density. Due to its small allowed range, this quantity is poorly constrained. The DTS ratio is an analogue of parameters that describe the differences between the dust and stellar disk in more general geometry. Our results imply the difficulty of constraining such parameters using SED fitting.
    \item The bulge radius and intrinsic bulge to total luminosity ratio have similar but weak effects on the SED. Hence, these parameters are unconstrained.
    \item We find that the inclination angle cannot be constrained due to the degeneracy in UV to optical wavelengths. If the inclination angle is given, the uncertainties in the inferred $A_\text{UV}$ and $A_\text{V}$ can be reduced by $\sim 0.1$ mag, indicating that the inclination correction is essential for studies that use SED fitting to measure dust attenuation curves.
    \item We study the impact of S/N on the SED fitting using a range of S/N values from 5 to 80. We find that high S/N data does not lead to better constraints on the geometry parameters. In particular, the improvements in the inferred SFH and galaxy parameters are limited when the S/N is increased from 20 to 80.
    \item We suggest that our SED model can be combined with image fitting tools to obtain better estimations on SFH and galaxy parameters. As an upper limit, we study the case where all the unconstrained parameters i.e. the inclination angle, DTS ratio, bulge radius, and intrinsic bulge to total luminosity ratio are fixed. The results indicate that the errors in the inferred disk radius, dust mass and dust attenuation can be reduced significantly. However, no improvements are found in the inferred SFH, stellar mass, SFR, and metallicity. 
\end{itemize}

\section*{Acknowledgements}
This work is partly supported by the NSFC (No.11825303, 11861131006), the science research grants from the China Manned Space project with NO.CMS-CSST-2021-A03, CMS-CSST-2021-B01, the Fundamental Research Fund for Chinese Central Universities (226-2022-00216), and the cosmology simulation database (CSD) in the National Basic Science Data Center (NBSDC-DB-10). We thank the reviewer for providing a constructive report, which improved the quality of the paper.
 \par
 In addition to those already been mentioned in the paper, we acknowledge the use of the following software: \textsc{astropy} \footnote{https://www.astropy.org/} \citep{astropy:2013, astropy:2018}, \textsc{jupyter-notebook} \footnote{https://github.com/executablebooks/jupyter-book}, \textsc{matplotlib} \citep{2007CSE.....9...90H}, \textsc{numpy} \citep{harris2020array}, \textsc{scipy} \citep{2020SciPy-NMeth}.

\section*{Data Availability}
The data underlying this paper will be shared on reasonable request to the corresponding author.




\bibliographystyle{mnras}
\bibliography{references}


\bsp	
\label{lastpage}
\end{document}